\documentclass[letterpaper,twocolumn,10pt]{article}
\usepackage{usenix}

\usepackage{tikz}
\usepackage{amsmath}
\usepackage[normalem]{ulem}
\usepackage{pifont}
\usepackage{array}
\usepackage{multirow}
\usepackage{xcolor}
\usepackage{listings}
\usepackage{booktabs}
\usepackage{tabularx}
\usepackage{csquotes}
\usepackage{quoting}  
\usepackage{enumitem}

\usepackage{graphicx}      
\usepackage{subcaption}    
\usepackage{amsmath}       
\usepackage{comment}
\usepackage{adjustbox}
\usepackage{caption}   
\usepackage{xcolor}

\definecolor{myTypeColor}{RGB}{200, 0, 0} 
\definecolor{myMacroColor}{RGB}{0, 0, 200} 

\newcommand{\mType}[1]{\textcolor{myTypeColor}{\textbf{#1}}}
\newcommand{\mMacro}[1]{\textcolor{myMacroColor}{#1}}

\usepackage{mdframed}
\definecolor{insightLineColor}{HTML}{45B7D1}

\newmdenv[
  topline=false,
  bottomline=false,
  rightline=false,
  linecolor=insightLineColor, 
  linewidth=3pt,     
  backgroundcolor=gray!5, 
  innertopmargin=5pt,
  innerbottommargin=5pt
]{sidebarinsight}

\definecolor{bg}{rgb}{0.98,0.98,0.98}
\definecolor{keywordcolor}{rgb}{0.6,0,0}
\definecolor{commentcolor}{rgb}{0.0,0.4,0.6}
\definecolor{stringcolor}{rgb}{0.0,0.5,0.0}
\definecolor{numbercolor}{gray}{0.5}
\quotingsetup{leftmargin=1.0cm, rightmargin=1.0cm}  

\setlist{itemsep=0pt, topsep=2pt, parsep=0pt, partopsep=0pt}
\begin{document}


\date{}

\title{\Large \bf GMM: Generalized OS Memory Management for Programmable XPUs}

\author{
Weixi Zhu\footnotemark[1]\\
weixi.zhu@openeuler.sh
\and
Bin Wang\footnotemark[1]\\
bw373717@gmail.com
\and
Zhiyuan Fang\footnotemark[2]\\
fangzhy27@mail2.sysu.edu.cn
\and
Jiahua Wang\footnotemark[3]\\
jiahuawang@stu.pku.edu.cn
\and
Yao Guo\footnotemark[3]\\
yaoguo@pku.edu.cn
\and
Gingfung Yeung\footnotemark[1]\\
ygingfung@gmail.com
\and
Adam Barker\footnotemark[1]\\
adam.barker@gmail.com
\and
Xinwei Hu\footnotemark[1]\\
shinwell\_hu@openeuler.sh
} 

\maketitle
\begingroup
\renewcommand{\thefootnote}{\fnsymbol{footnote}}
\footnotetext[1]{openEuler community}
\footnotetext[2]{Sun Yat-sen University}
\footnotetext[3]{Peking University}
\endgroup

\begin{abstract}





Since the AI boom, the fast evolving programmable XPU devices utilize 
high-bandwidth local 
memory to accelerate domain-specific tasks offloaded from the host CPU,
resulting in a proliferation of XPU memory management systems implemented
in proprietary drivers. 
However, existing OS core memory management (MM) continues to lack a 
high-level software abstraction, excluding the XPUs from leveraging 
OS advanced memory features. The XPU manufacturers have to reimplement 
independent XPU MM drivers, exacerbating the fragmented XPU software 
ecosystem.

This paper presents GMM, generalized memory management that provides unified MM subsystems for XPUs. By adding an explicit logical mapping abstraction, GMM
1) provides universal NUMA-aware memory API for XPUs
2) allows ease of XPU MM delegation
3) enables sharing of advanced XPU memory features
Lightweight XPU MM drivers are implemented for two representative 
XPU manufacturers, Nvidia and Huawei.
Both GMM-based drivers obtain competitive performance compared with their vendor-supplied drivers. GMM also obtains up to 41\% higher fault-driven page migration throughput. 
Two real-world industrial case studies are exercised, where GMM reduces 
XPU memory utilization by 10\% with up to 22.6x faster inference serving switch latency.

\end{abstract}

\section{Introduction}

\begin{figure}[t]
    \centering
    \includegraphics[width=0.45\textwidth]{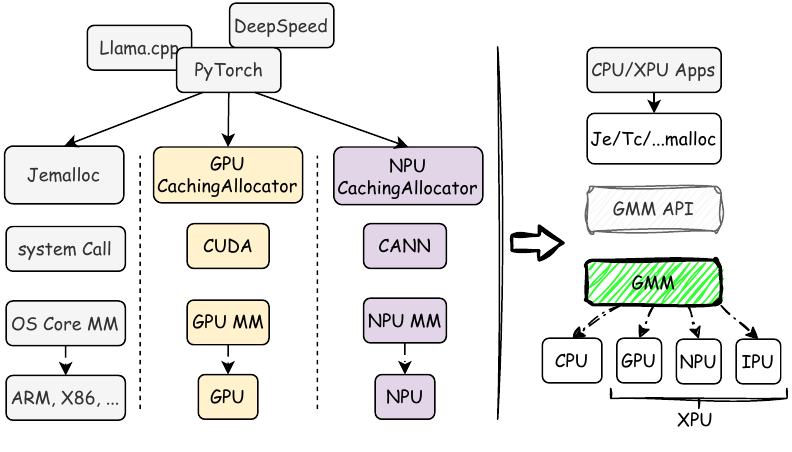}
    \caption{GMM unifies the fragmented XPU memory ecosystem}
    \label{fig:gmm_intro}
\end{figure}

Programmable domain-specific accelerators (XPUs) such as GPUs, NPUs 
have become a central part of modern datacenters~\cite{
norrie2021design, jouppi2017datacenter, nickolls2010gpu, 
dally2020domain, kaplan2024intel, rico2024amd, smith2025amd, 
liao2021ascend}. 
These devices execute computation on high-bandwidth local memory and 
expose rich virtual memory capabilities, including multi-level page 
tables and hardware page faults~\cite{emt-ref62, emt-ref69, emt-ref81
, emt-ref84, zhu2020comprehensive}. 
Despite this convergence with CPU memory systems, today's operating 
systems still exclude XPUs from the core memory management (MM) 
subsystem. XPU vendors must ship large proprietary 
memory managers within their drivers~\cite{nvidia_rm, amd_rocm, 
intel_xe}. 
Tens of thousands of lines of code must be re-implemented to include 
core MM features such as paging, defragmentation and 
oversubscription, leading to fragmented XPU memory software as shown 
in Figure~\ref{fig:gmm_intro}. 

As modern datacenters become increasingly heterogeneous and contain 
multiple XPU manufacturers (e.g., Nvidia, Huawei) across their 
fleets, this fragmentation has immediate practical consequences. 
\textit{First}, each vendor reinvents similar MM functionality, 
forcing frameworks (e.g., PyTorch~\cite{paszke2019pytorch}, 
llama.cpp~\cite{llamacpp}) to maintain multiple 
backends with inconsistent semantics and underlying performance. 
\textit{Second}, XPUs cannot easily leverage decades of OS MM 
innovations~\cite{cortenmm-ref9, cortenmm-ref10, zhang2025cortenmm, chai2025emt}, such as page laundering, NUMA policies~\cite{emt-ref37, emt-ref61, emt-ref71, emt-ref92, emt-ref78}, memory hints 
etc., because these features are deeply entangled with CPU-centric 
assumptions. 
\textit{Third}, coordination between two memory management systems 
introduces performance issues, as they lack control-plan information 
from each other.
As a result, deep learning frameworks start to reinvent part of the 
memory subsystems to compensate for missing MM support~\cite{
devito_cuda_caching_2022, guo2024gmlake, peng2020capuchin, 
wang2018superneurons}.

Many XPU manufacturers implement eager memory management systems that 
lack the flexibility of allocating physical memory on demand. This 
puts heavy portability burdens on shoulders of application 
developers, who need to carefully manage XPU memory utilization 
among different generations of XPU products of different XPU 
manufacturers. Furthermore, the limited XPU local 
memory gets easily fragmented due to the eager allocation semantic~\cite{guo2024gmlake}, 
leading to ad-hoc reinvention of \texttt{malloc()} libraries for XPU 
applications.

Linux HMM encodes CPU page tables to store XPU status, providing 
event-driven hooks to redirect page fault handling for XPU MM 
drivers. This enables the XPU driver to support unified virtual 
memory by handling both CPU and XPU memory management~\cite{
allen2021depth, shao2022oversubscribing, ganguly2019interplay, 
choi2022memory, allen2024fine, long2023deep, pratheek2024suv, 
zhu2022uvm}, but introduces significant performance degradations
by abusing CPU arch-related data structures to orchestrate MM
coordination.
Furthermore, HMM still lacks a software abstraction
that allows the XPU MM to leverage core MM's advanced memory
features. The XPU driver still needs to implement an independent
memory management subsystem.
Additionally, the HMM coordination comes at a price, because 
the XPU MM must correctly maintain extra OS kernel data structures 
following the core MM pattern.

Therefore, the key missing piece is an abstraction that allows: (1) 
XPUs to be coordinated alongside CPUs within the core MM; (2) While 
preserving each devices native page table format, page size and 
performance properties; (3) Without entangling vendor-specific code 
within the OS kernel. No such abstraction exists today.

\noindent \textbf{Contributions.} This paper introduces GMM, generalized
memory management that provides unified MM subsystems for XPUs. GMM 
elevates programmable XPUs to first-class MM participants. 

Specifically, GMM is centered around the logical mapping abstraction which tracks 
the physical data location across CPU and XPUs. GMM then extends the 
generic page fault handler to handle XPU mappings invoking MMU 
functions registered by the aforementioned GMM KPI. Additional 
advanced memory features including coherent view of memory, XPU 
memory oversubscription, XPU memory hints are included to be 
leveraged by XPUs.

GMM redesigns the core MM to provide NUMA-aware memory API. The 
vendor-agnostic API enables unified pointers for both portability 
and programmability. Additionally, a high-level kernel programming 
interface (KPI) is provided for XPU drivers to delegate XPU memory 
management to the core MM, erasing the demand of reinventing any 
independent memory management in the XPU driver.

Lastly, GMM adopts simple yet efficient design choices based on 
insights of XPU application and hardware commonalities to learn the 
performance implications of sharing the core MM. For example, 2MB 
page queues are maintained for XPU physical memory management which 
brings competitive XPU memory performance and data transfer 
throughput.

The result is that XPUs can leverage the OS's existing MM 
implementations, e.g. page fault handling and memory oversubscription. 
Lightweight GMM-based drivers are implemented with only a small set of MMU 
operators (~500 - 2000 LoC) replacing tens of thousands of lines of 
bespoke driver code. These lightweight drivers also demonstrate 
competitive end-to-end performance, with up to 41\% higher fault-driven 
page migration throughput. 
Case studies stemmed from industrial scenarios showed that GMM 
brings up to 22.6x faster model switch latency for LLM inference 
serving, as well as reduces 10\% memory utilization enabling a universal \texttt{jemalloc} library for XPUs.

In summary, this paper makes the following contributions: 

\begin{itemize}
    \item We propose GMM, a straightforward OS core MM design that allows XPUs to leverage advanced OS memory features.
    \item We present thorough comparison and evaluation of the existing independent XPU MM drivers.
    \item We demonstrate practical experience of implementing lightweight GMM-based XPU drivers for two representative vendors with their performance implications.
    \item We derive insights for existing XPU MMU characteristics and future hardware design directions from the perspective of a unified MM.
\end{itemize}

\section{Background}
\label{sec:background}

\subsection{Virtual Memory Management}


The memory management system provides a virtual memory abstraction
for ease of programmability and security isolation at the 
cost of virtual-to-physical address translation. Commercial 
processors, including CPUs and XPUs, typically employ a radix 
tree-based page table~\cite{emt-ref11, emt-ref17, emt-ref19, nvidia-gp100-mmu}. The page table is indexed by different bits 
of the virtual address for each hierarchy, pointing to a single page 
table entry describing a linear mapping relationship between the 
indexed virtual page and the mapped physical page. The processor's 
memory management unit (MMU) then utilizes a translation look-aside 
buffer to cache the page table entries for fast look-ups of the 
address translation.

Memory management systems of existing CPUs or XPUs are designed with 
similar software architectures.
When a process starts, a virtual address space must be created to 
isolate its code and data from other processes. To manipulate data, 
virtual addresses must be allocated. The memory system can defer the 
time of physical memory allocation and page table entry installation 
until a page fault gets triggered, indicating memory accesses to a 
virtual address. Some memory systems choose to eagerly fault the 
memory at the allocation time, avoiding future page faults under the 
risk of increased memory fragmentation. 
At the time, most CPU memory systems (OS core MM) defers physical 
memory allocation to serve generalized applications~\cite{
cortenmm-ref9, cortenmm-ref12, cortenmm-ref16, cortenmm-ref43}, 
while XPU 
memory systems usually choose a eager allocation design 
decision~\cite{nvidia_rm, amd_rocm, intel_xe}.

The page table format is defined by hardware manufactures, demanding 
coordination from the memory management systems. 
Therefore, these systems usually implement a hardware-agnostic layer 
to hide the differences of page table manipulations. 
Linux core MM assumes that the page table can be directly 
manipulated, so it implements a multi-level radix tree-based page 
table abstraction manipulated by low-level page table functions. 
However, XPU MMs are different. 
Each XPU MM is bound with a specific manufacturer, 
and the driver code running on the CPU cannot directly access the 
XPU local memory. 
Therefore, they implement and isolate vendor-specific XPU page table 
manipulation command for different generations of their products.

\subsection{Unified Virtual Address Space}

There is a rising demand of unified virtual address space~\cite{
shao2022oversubscribing, long2023deep, pratheek2024suv, zhu2022uvm}, 
because
a XPU application usually involves both CPUs and XPUs which solve 
complementary tasks. Consequently, programmers must allocate 
pointers from two independent virtual address spaces and manually 
track which pair of pointers logically refers to the same data. 
When both the CPU and XPU processors operate on the same data, 
programmers need to synchronize the computation with the data transfers correctly.

Nvidia's UVM driver implements such unified virtual memory support.
The driver exploits Linux's file page fault mechanism, which 
redirects core MM page fault handling to the device file abstraction 
that represents the GPU device. 
Therefore, the UVM driver takes control of both the GPU page faults
and CPU page faults, creating a coherent view of memory by transparently
migrating physical pages.

Linux also provides heterogeneous memory management (HMM) for XPUs. 
However, unlike GMM, HMM does not allow XPU MMs to leverage the core 
MM memory features or delegate memory management to the core MM.
The XPU MM still needs to implement a full memory management system 
and utilize HMM's interface to interact with the core MM.
HMM provides an event-triggered interaction mechanism which tracks 
the device page table status within CPU page table entries,
so the core MM can ask the XPU MM to migrate back a device page 
upon a CPU page fault.  
This mechanism allows XPU MM drivers to interact with anonymous 
memory instead of exploiting the file page fault mechanism. 
As a result, XPU memory is allocated via the device-agnostic core MM 
interface \texttt{mmap()} instead of driver's own \texttt{ioctl()}.

Some hardware solutions employ a cache-coherent interconnect 
between the CPU and XPUs to let the core MM directly manage a NUMA 
node. 
This does not waive the effort of improving the existing core MM, 
which is unaware of the heterogeneity. For example, the unmodified core
MM will map a DDR page for the CPU process which triggers a XPU page fault, 
leading to severe throughput degradation. 
Furthermore, such solution is limited to the case where both CPU and 
XPU manufactures integrate the bus protocol of the interconnect, 
which limits the combinations of CPU and XPU hardware manufactures.

\section{The need for a Generalized MM}
\label{sec:motivation}


\begin{table}[t]
\begin{tabular}{cccccc}
\toprule
Linux & FreeBSD & NV-RM & UVM  & ROCm & Xe    \\ \midrule
141 K & 66 K    & 26 K  & 69 K & 27 K & 11 K  \\ \bottomrule
\end{tabular}
\caption{\#LoC Estimations of CPU/XPU MM systems}
\label{table:mm_loc}
\end{table}

\subsection{Fragmented XPU Software Ecosystem}

The OS core MM enables the CPU applications to use a universal API 
to program different CPU ISAs, but each XPU manufacturer requires a 
different vendor-specific API. 
The root cause is the OS core MM lacks a hardware-agnostic 
abstraction to handle XPUs, so each XPU reinvents a similar MM 
system in its driver. Such MM systems are complicated and costly,
as approximated in Table~\ref{table:mm_loc}.
The fragmented XPU MM systems further lead to fragmented 
application-level support. 
For example, deep learning frameworks like PyTorch need to implement 
different memory allocation backends for different XPU manufactures~\cite{
dlframework1, dlframework2, dlframework3, dlframework4, dlframework5, dlframework6, dlframework7, dlframework8, dlframework9, dlframework10, dlframework11}. 
As XPU manufactures and applications grow, the combination 
ultimately lead to a problem of reinventing existing solutions.

\begin{figure}[t]
    \centering
    \includegraphics[width=0.45\textwidth]{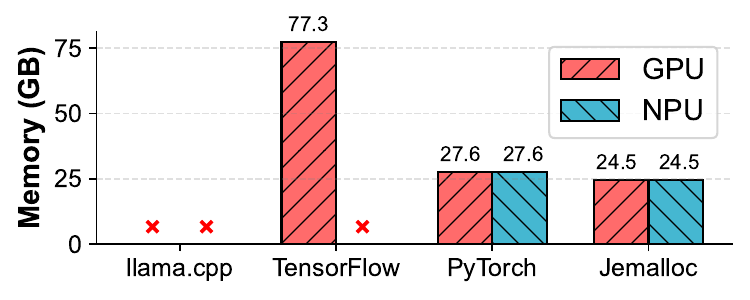}
    \caption{Memory consumption running a memory trace for 
        OPT-1.3B training. Some combinations of frameworks and XPUs 
        do not support dynamic memory allocation.}
    \label{fig:memory_allocate}
\end{figure}

\begin{figure}[t]
    \centering
    \includegraphics[width=0.45\textwidth]{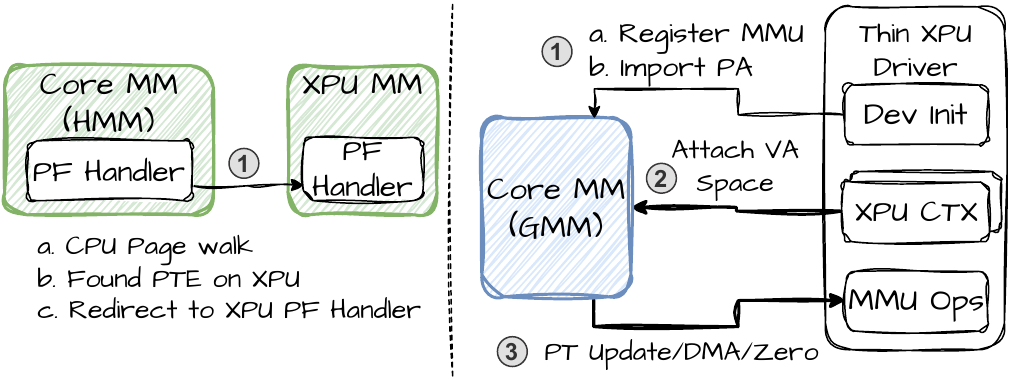}
    \caption{Coordination comparison between HMM and GMM.}
    \label{fig:gmm_vs_hmm}
\end{figure}

\begin{figure}[t]
    \centering
    \includegraphics[width=0.45\textwidth]{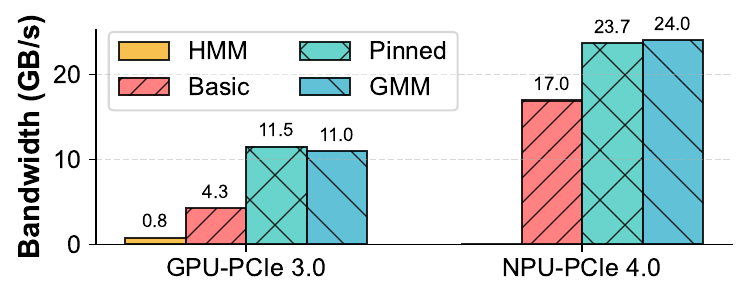}
    \caption{Bandwidth of buffer copying between the CPU and XPU. Basic copies between buffers allocated by XPU MM and core MM. Pinned allocates pinned CPU device memory with XPU MM. GMM allocates unified buffers by core MM.}
    \label{fig:bandwidth_comparison}
\end{figure}

\begin{figure}[t]
    \centering
    \includegraphics[width=0.45\textwidth]{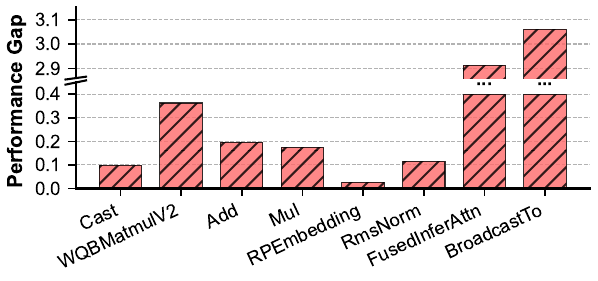}
    \caption{Performance gap using 4KB over 2MB mappings on Qwen-2.5-3B prefill stage over top-8 frequent NPU kernels (sum up to 92.6\% kernel invocations). Using 2MB mappings brings 35\% faster prefill speed.}
    \label{fig:operator_gap_cdf_analysis}
\end{figure}

As XPUs have limited memory capacities, memory fragmentation becomes 
a major problem in the XPU infrastructure. 
The memory fragmentation is 
mostly because XPU MMs implement eager memory allocations. 
Recent work explores application-level defragmentation mechanisms in 
the XPU frameworks~\cite{wang2018superneurons, shriram2019dynamic, peng2020capuchin, guo2024gmlake, devito_cuda_caching_2022}, 
or even reinvents a shadow page table mechanism inside the XPU 
kernel named PagedAttention~\cite{dlframework4}, 
but they are ultimately reinventing well-studied mechanisms, 
e.g. the \texttt{malloc()} libraries~\cite{vAttention, dice2002mostly, evans2006scalable, hunter2021beyond, hudson2006mcrt, leijen2019mimalloc, yang2023numalloc, tcmalloc_overview}. 
Figure~\ref{fig:memory_allocate} exercises XPU memory allocators 
with a memory trace of training OPT-1.3B. It shows that PyTorch's
CachingAllocator is effectively a poor reinvention of Jemalloc, 
while TensorFlow 
does not support 910B NPUs and llama.cpp does not support dynamic 
memory allocations for both XPUs.

\begin{sidebarinsight}
    \textbf{Insight 1: The XPU software fragmentation is exacerbating, demanding a generalized XPU MM design to leverage existing work.}
\end{sidebarinsight}

\subsection{Difficulties of MM Coordination}

Coordination between two complicated MM systems can be difficult. 
This can be exemplified with data transfers between the CPU and the 
XPU.
Figure~\ref{fig:bandwidth_comparison} compares the data transfer 
bandwidth between two buffers allocated for the CPU and the XPU.
``Basic" refers to allocating XPU memory with vendor-specific APIs 
and CPU memory with \texttt{malloc()}, while ``Pinned" refers to 
allocating both CPU and XPU memory with vendor-specific APIs, where 
the CPU memory is mapped to pinned device kernel pages. 
GMM allocates unified pointers with OS core MM for both CPU and XPU. 
Both XPUs exhibit noticeable bandwidth advantages over ``Basic", 
which requires coordination between two independent MM systems.

\begin{sidebarinsight}
    \textbf{Insight 2: Using a single generalized MM is strictly simpler and more efficient than coordinating two autonomous MMs.}
\end{sidebarinsight}

\subsection{Limitations of Existing Work}

Nvidia's UVM driver implements a one-MM solution for CPU-GPU unified 
virtual memory. Memory allocated by the UVM API can be used by both 
the CPU and the GPU. This is achieved with Linux's file-backed 
memory, which uses VMA operators to redirect CPU page faults to the 
device driver. 
As a result, it is a one-MM solution that only works for Nvidia's 
GPU with Nvidia's UVM API.

Figure~\ref{fig:gmm_vs_hmm} compares HMM's coordination pattern 
with GMM. 
Linux HMM is designed to allow a XPU MM to coordinate with the CPU 
MM, enabling support for unified virtual address space for XPU 
programmability. 
This is achieved by encoding the CPU page table entries to represent 
XPU mapping status, along with callback functions to redirect the 
CPU page fault to the XPU MM over a PTE mapped on the CPU. 
This allows hardware-agnostic Linux's anonymous memory to be 
utilized over XPUs, but effectively provides similar coordination 
mechanism as file-backed memory. 
Furthermore, abusing the CPU PTEs mandates XPU 
page sizes of 4KB. This can cause significant overhead for PCIe transfers, as well as kernel computation, as shown in 
Figure~\ref{fig:bandwidth_comparison} and Figure~\ref{fig:operator_gap_cdf_analysis}.

\begin{sidebarinsight}
    \textbf{Insight 3: No existing work has tried to extend core MM for XPUs, but only increases the coordination mechanism.}
\end{sidebarinsight}
\section{GMM Design}
\label{sec:design}

\begin{figure}[t]
    \centering
    \includegraphics[width=0.45\textwidth]{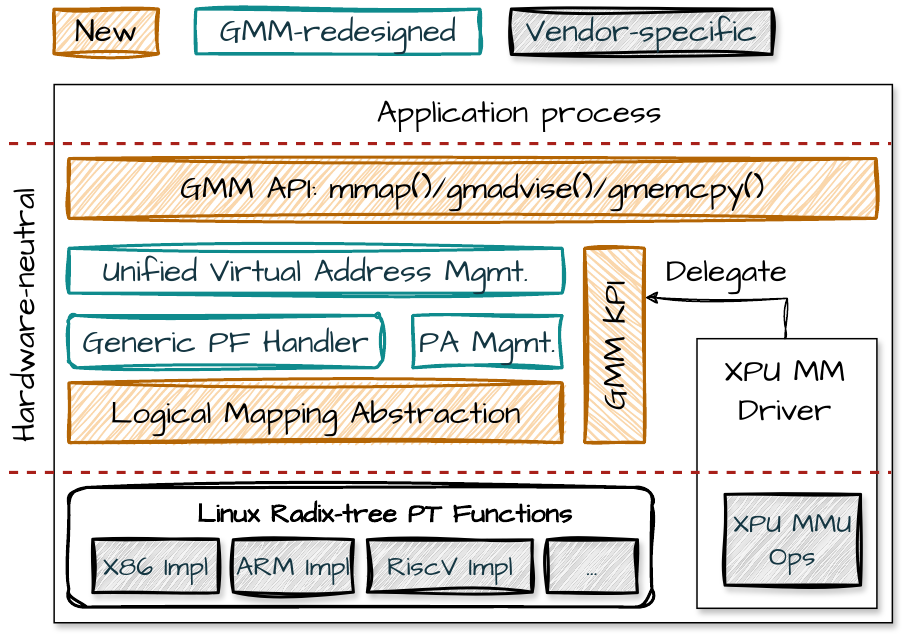}
    \caption{Overview of GMM architecture}
    \label{fig:gmm_overview}
\end{figure}

GMM's core design principle is KISS -- keep it simple and stupid. 
GMM makes simple design decisions based on XPU commonalities and 
introduces no complicated performance optimizations.
The goal of GMM design is to help the empirical evaluation learn clear
performance implications of leveraging core MM features.

Figure~\ref{fig:gmm_overview} illustrates a high-level overview of GMM.
GMM is centered around an explicit logical mapping abstraction.
Application processes utilize GMM API to allocate virtual memory 
within unified virtual address spaces. The API provides NUMA-aware 
interface for data locality optimizations. 
A kernel programming interface (KPI) is also provided for XPU 
drivers to delegate memory management, which helps XPU driver import 
the XPU physical address range, register MMU functions and attach 
XPU context to CPU address space.
GMM then redesigns the generic page fault handler to look up the 
logical mapping abstraction and resolve XPU page faults with XPU MMU 
functions, orchestrating page zero-filling or migration. 
A few detailed design goals are listed as follows:

\begin{itemize}
\item \textbf{A universal NUMA-aware memory API.} 
GMM extends system calls to allocate CPU-XPU unified pointers for
XPU programmability. NUMA-aware API further allows performance
optimizations with data locality preferences.

\item \textbf{Ease of XPU MM delegation.}
GMM is centered around the logical mapping abstraction. XPU drivers
can delegate XPU memory management to core MM with GMM's kernel
programming interface by providing MMU operation functions.

\item \textbf{Advanced XPU memory features.} 
The core MM is extended to support advanced memory features
leveraged by GMM for XPUs, including a generic page fault
handler for XPUs, XPU memory oversubscription and XPU-agnostic
\texttt{malloc()} libraries.

\item \textbf{Negligible overhead for CPU workloads.} 
GMM's logical mapping abstraction skips tracking CPU-only 
memory allocations, so Linux's CPU page table optimizations
can remain effective for CPU-only workloads.
\end{itemize}

\subsection{GMM Interface}
\label{subsec:interface}

\begin{table}[t]
\adjustbox{width=\columnwidth}{
    \begin{tabular}{ll}
    \toprule
    Interface                          & Description               \\ 
    \midrule
    mmap(MAP\_XPU\_VISIBLE)            & Allocate unified pointers \\
    gmadvise(numa\_id, va, size, hint) & NUMA-aware memory hints   \\
    gmemcpy(numa\_id, dst, src, size)  & Buffer copy with NUMA destination   \\ 
    \bottomrule
    \end{tabular}
}
\caption{GMM API for User-level Programming.}
\label{tab:gmm_api}
\end{table}

\begin{table}[t]
\adjustbox{width=\columnwidth}{
\begin{tabular}{lll}
\toprule
        & Interface                  & Description                      \\ \midrule
gm\_dev & \_create(\&mmu\_ops, ...)  & Register XPU MMU ops             \\
        & \_import(dev, pa, size)    & Delegate XPU PA Management       \\
        & \_fault(dev, va, rw)       & Delegate XPU Page fault          \\ \midrule
gm\_mmu & (va\_examine)(va, size)    & Examine XPU VA availability      \\
        & (pmap\_enter)(va, pa, ...) & Create XPU page table entries    \\
        & (pmap\_clear)(va, size)    & Destroy XPU page table entries   \\
        & (dma\_copy)(dst, src, ...) & Issue a DMA copy command         \\
        & (zero)(va, size)           & Zero XPU memory                  \\ \midrule
gm\_as  & \_create(gm\_as, ...)      & Register XPU context             \\
        & \_attach(gm\_as)           & Attach XPU context to current mm \\
        & \_destroy(gm\_as)          & Destroy XPU context              \\ \bottomrule
\end{tabular}
}
\caption{GMM KPI for XPU MM driver programming. 
Only hardware-specific functions should be implemented.}
\label{tab:gmm_kpi}
\end{table}

\paragraph{Unified NUMA-aware Memory API.}

Table~\ref{tab:gmm_api} lists GMM's unified NUMA-aware memory API 
for XPU applications. The API enables programmers
to quickly implement CPU-XPU applications with unified pointers that 
are XPU-agnostic, with NUMA-aware APIs for data locality 
optimizations.

To allocate memory, the CPU program should first initialize the XPU 
device context. It then invokes \texttt{mmap()} system calls with 
\texttt{MAP\_XPU\_VISIBLE} flag set. The returned pointers are 
unified between CPU and the initialized XPU.
Either processor can expect reading or writing
coherent values via the same virtual address regardless of the 
underlying CPU-XPU interconnects. 
An \texttt{munmap()} call should be coupled to deallocate memory.

Two system calls are added to provide user-level NUMA-related 
performance optimizations. \texttt{Gmadvise()} adds one argument 
denoting a NUMA node to operate memory hints on unified virtual 
address ranges. \texttt{Gmemcpy()} executes data transfers between 
two unified pointers, guaranteeing that the destination buffers will 
end up being mapped on the designated NUMA node.

\paragraph{XPU MM Delegation.}

Table~\ref{tab:gmm_kpi} lists GMM KPI that provides high-level kernel
programming interface for XPU drivers to delegate XPU
memory management to the core MM. GMM KPI has three sets of GMM KPI 
for XPU device initialization, XPU-specific MMU functions, and address 
space association.

XPU Device initialization includes three steps. First, a device 
instance is created with vendor-specific implementation of MMU 
functions registered. The device instance then gets a unique NUMA node
identifier representing the XPU local memory. Last, a range of XPU 
physical address is imported for GMM to allocate. 
Upon XPU page fault interrupt, the XPU driver should invoke 
\texttt{gm\_dev\_fault()}
which wraps core MM's generic page fault handler.

The MMU operators include five device-specific functions. 
\texttt{va\_examine()} is invoked by GMM at the \texttt{mmap()} time
for XPU drivers to validate whether the allocated virtual addresses 
are legal. 
\texttt{pmap\_enter()} installs virtual-to-physical linear mappings
in the XPU page table given proper XPU mapping protections.
\texttt{pmap\_clear()} revokes the mapping of a given virtual address
range, and clears the XPU MMU TLB if applicable. 
\texttt{Memcpy()} and \texttt{zero()} manipulate XPU physical data, 
orchestrating DMA copying or zero-filling operations.

Each XPU context must be associated with the corresponding CPU process
via the \texttt{gm\_as} interface. The XPU context is created before
the CPU program can offload computation tasks to the XPU. This is
analogous as a CPU \texttt{fork()} call, where the
XPU page table should be created by the XPU driver. Additionally,
the XPU context should be attached to the address space of the
CPU process, maintained as unified virtual address space by GMM.

\subsection{Unified Virtual Address Space}

GMM provides unified virtual address space (UVAS) at the buffer level.
This avoids making strong assumptions that any addresses allocated 
by the CPU process can be available in the XPU context. 
Therefore, GMM only guarantees that virtual addresses returned by 
\texttt{mmap(MAP\_XPU\_VISIBLE)} are valid on both CPU and XPUs. 
GMM implements an examine-and-retry loop to valid if the virtual 
addresses allocated by the core MM are XPU-available. 
The examination invokes \texttt{va\_examine()} provided by the XPU 
MMU operators. 
Each retry loop calls the internal VMA allocation function to 
reserve a VA range from the CPU virtual address space. 
The reserved VMAs are freed until the XPU VA examination passes or 
the retry time reaches the limit.

\subsection{Logical Mapping Abstraction}
GMM reuses Linux's radix tree data structure, named \texttt{xarray} in 
Linux, to index the entries of logical mapping information of each 
virtual page 
allocated by \texttt{mmap(MAP\_XPU\_VISIBLE)}. Linux's \texttt{xarray} 
implementation is cache-optimized and RCU-friendly.
Each mapping entry indicates whether the indexed 2MB 
virtual page is already backed by a physical 2MB page.

The logical mapping abstraction can slightly increase the 
concurrency overhead of the core MM. It has a short critical section 
at the entry allocation and deallocation time. The 
allocation happens at the first page fault time, and will not be 
deallocated until \texttt{munmap()}, or the process exits. 
However, the entry lookup is lock-free, and each entry holds a spin
lock for fine-grained concurrency control of page table operations
for both CPU and XPUs. This does not introduce any additional concurrency 
bottleneck since Linux uses a PMD-level page table lock.

\paragraph{Coherent view of memory.}

The logical mapping abstraction allows implementing a coherent view
of memory within the UVAS. Given a sequence of the CPU and XPU memory
accesses to the same virtual address, each processor is guaranteed to
operate on the exact updated data after the last memory access.
GMM can maintain the coherence with two methods.
One is to keep a single hardware mapping for each virtual page,
so any subsequent memory accesses will trigger a page fault
which migrates data. The other way is to leverage cache-coherent links
between the CPU and XPU, and create a hardware mapping for remote 
physical page.

GMM prioritizes XPU data locality to suit the significant 
memory bandwidth of XPUs. This avoids relying on the cache-coherent
interconnects that may limit the combination of CPU and XPU 
manufactures. However, GMM can still create remote mappings with
MMU functions.

\subsection{Simplified XPU Page Management.}

GMM implements physical XPU page lists for a single page size support
of 2MB. The reason is because XPUs do run general-purpose
kernel code which require different orders of pages for device drivers,
so page lists suffice. Besides, the 2MB page size is the default
choice for XPU manufactures, which is also widely supported by
CPU manufactures.

The page lists describe two page status --- active and free. 
The active page list is maintained as an LRU-ordered FIFO queue, 
indicating a history of the page hotness. 
Any event that tries to map a physical page will dequeue and enqueue
it, reinserting it to the most recently used side.
Such a event can be a page fault or a memory hint of \texttt{MADV\_WILLNEED}.

\paragraph{XPU page reclamation.}

The XPU memory can be oversubscribed by automatically evicting XPU 
pages to the host DDR, analogous to CPU memory oversubscription by
swapping DDR pages to a swap disk.
The eviction process is synchronously triggered upon a XPU page 
allocation failure. Sixteen victim 2MB XPU pages are dequeued from
the XPU active page queue.
Their reverse mapping information is maintained in the XPU page data
structures to help
GMM invoke the XPU MMU operator to clear the XPU page table entries and 
transfer their data back to the host DDR.
The XPU pages then get freed and enqueued to the free page list. 
The logical mapping abstraction gets updated
to track the CPU physical pages which hold the evicted data. 
However, no CPU PTEs get prematurely created to map the evicted CPU 
pages, as future memory accesses are expected to come from XPUs. If 
the CPU re-accesses evicted pages, then CPU page faults are triggered
to create CPU PTEs.

\subsection{Generic Page Fault Handler}

GMM extends the core MM page fault handler to carry an extra unique
processor identifier which represents either the CPU or a XPU. 
The page fault handler then allocates physical pages with a simple 
first-touch NUMA policy.
For the CPU, GMM picks the NUMA node affiliated with the CPU core 
that runs the faulted CPU process. 
For XPUs, GMM assumes a one-to-one NUMA mapping relationship, but it 
can be extended if XPUs evolve to have internal NUMA effects.

When a page fault gets triggered, GMM first examines whether the 
faulted virtual address belongs to a Linux VMA allocated with 
\texttt{MAP\_XPU\_VISIBLE} set. 
If not set, then GMM confirms the faulted processor is a CPU and 
skips the interactions with logical mapping abstraction. 
The subsequent page fault procedure executes the performant core-MM 
code dedicated for CPU-specific memory, which uses hardwired 
manipulation functions over CPU page tables. 

If \texttt{MAP\_XPU\_VISIBLE} is set, GMM looks up the logical 
mapping entry of the faulted virtual address and allocates a 
2MB physical page from the NUMA node of the faulted processor 
identifier. 

If the entry is empty, then GMM acquires the xarray's giant lock to 
allocate and initialize the entry.
If the entry does not point to any backed physical page, GMM 
zero-fills the physical page. 
For CPU processor, GMM borrows the existing page zero-fill routine. 
For XPU processor, GMM invokes the XPU MMU function \texttt{zero()}. 
GMM then holds the writer lock of the entry to 
1) update the entry to track the allocated physical page. 
2) update the processor page table with CPU or XPU MMU function 
\texttt{pmap\_enter()}. 
The XPU page table is thus modified under fine-grained lock 
protection at the 2MB virtual page level. 
This is a similar protection protocol as Linux uses PMD-level lock 
on CPU page table manipulations.

If the backing physical page exists, GMM's behavior divides into two 
branches. 
In case the backing physical page already resides on the faulted 
processor, GMM only needs to install a hardware mapping to the 
resident physical page with the appropriate MMU function, and the 
previously allocated physical page can be released. 
If the backing physical page resides on another processor, then GMM 
needs to uninstall the hardware mapping by looking up the reverse 
mapping and invalidate the corresponding TLB entries. 
GMM then transfers the data to the previously allocated physical 
page on the faulted processor and installs a hardware mapping to it. 
The logical mapping entry then gets updated simultaneously.

\paragraph{Page fault coalesce.}

\begin{figure}[t]
    \centering
    \begin{lstlisting}[xleftmargin=15pt]
|\mType{gm\_ret}| |\mMacro{gm\_as\_create}|(as, ...)
    as->last_pfn = NULL

|\mType{gm\_ret}| |\mMacro{gm\_dev\_fault}|(dev, va, rw)
    pfn = get_xpu_pfn(va)
    as = find_address_space(current->mm, dev)
    if as->last_pfn == pfn:
        return GM_BUSY
    else
        as->last_pfn = pfn
        ... 
\end{lstlisting}
    \caption{GMM's page fault coalescing.}
    \label{code:migrate}
\end{figure}


GMM implements a straightforward page fault coalescing mechanism with six lines of code, as
illustrated in Figure~\ref{code:migrate}. 
The mechanism book-keeps the last virtual page and skips the current 
XPU page fault if they are the same page.

\subsection{Memory Hints}

GMM extends Linux \texttt{madvise()} system call to carry a new 
argument identifying the NUMA node to support VMA-level memory hints 
for CPU-XPU programming. 
Two memory hints of GMM are exercised in this paper, 
including \texttt{MADV\_WILLNEED} and \texttt{MADV\_DONTNEED}.

A \texttt{gmadvise} call with \texttt{MADV\_WILLNEED} hints GMM that 
the designated virtual range will be utilized on the designated NUMA 
node in the near future. 
GMM will always book-keep the user's hint by moving the
physical pages to the most recently used side of the active page
queue to guide physical page reclamation priorities.
Besides, a data prefetching operation is executed.
GMM queries the logical mapping abstraction and iteratively invokes 
the internal routine of the generic page 
fault handler to migrate data to the destination NUMA node for pages
not resident on the designated NUMA ID. 
This allows the user program to optimize data locality, or overlap 
the data transfers with future computation in advance. 
Note that the concurrency of page faults allows the data transfers 
to happen simultaneously with the computation, 
so a strict copy-then-execute data dependency is not required to be 
guaranteed by the programmer. A \texttt{procfs} interface is also 
provided
for system-level administrators to invoke \texttt{MADV\_WILLNEED} to 
prefetch all mapped virtual
pages of a process to a specified NUMA node.

A \texttt{gmadvise} call with \texttt{MADV\_DONTNEED} hints GMM that 
the designated virtual range will not be utilized in the near 
future, and it expects eager destruction of the underlying hardware 
mappings to release physical memory for memory defragmentation. 
This memory hint is utilized internally by \texttt{malloc} libraries. 
GMM provides the same semantics of this memory hint for CPU-XPU 
unified pointers, by destroying both the logical mapping entries and 
calling MMU functions to destroy the XPU page table entries.
This enables us to implement \texttt{jemalloc} for AI applications 
with minimum effort.
\section{Implementation Details}
\label{sec:impl}

GMM is implemented based on Linux 6.6, costing 2,500 lines of 
code, including 450 lines for logical mapping, 250 lines for 
interface, 500 lines for XPU physical memory management including 
200 lines for XPU page reclamation, 350 lines for memory hints and 
memory copies, 500 lines for the generic page fault handler and 150 
lines for statistics and debugging.

\paragraph{Internally parallelized prefetching.}

GMM uses a maximum of four kernel workqueues to parallelize data 
prefetching for \texttt{MADV\_WILLNEED} memory hints. Data migration 
tasks are partitioned at 2MB page granularity.

\paragraph{Huawei NPU Integration.}

The GMM-based Huawei NPU MM driver was implemented with 2,000 lines 
of code, including a 850-line NPU-specific DMA message channel 
multiplexer. The multiplexer partitions a reserved DMA buffer to 
concurrently poll the status of control plane MMU functions. 
The MMU functions reuse low-level function code from the NPU
firmware, which implements page table functions for a trimmed 
version of ARM MMU.

\paragraph{Nvidia GPU Integration.}

The GMM-based Nvidia GPU MM driver costs 680 lines of code. The GPU 
MMU operation reuses high-level function code from the Nvidia UVM 
driver, including the GPU page table manipulation, DMA engine and 
etc. The GPU page fault interrupt entry function is modified to 
invoke GMM's page fault handler. We calculated the development effort, 
and the integration process took three weeks of a 
four-year experienced kernel engineer.

\paragraph{Application-level Integration.}

It is straightforward to utilize GMM API in C-like XPU 
programs. \texttt{Jemalloc()} 5.3.0 is integrated with GMM, 
modifying three lines of code. It is then integrated with multiple 
XPU application frameworks, including PyTorch, MindSpore, 
and llama.cpp. PyTorch and MindSpore cost 20 lines integrating 
\texttt{Jemalloc()} for XPUs and 100 lines replacing the XPU memcpy 
APIs. These numbers are 40 and 60 for llama.cpp integration, 
respectively.

\paragraph{Functional Correctness.}

A strict product quality process is applied to empirically guarantee
the functional correctness of GMM. The process includes a total of 
52K test cases, with 975 tests on the core MM system. 51 of these 
test cases are additionally included for GMM's advanced memory 
features. 

\section{MM Systems Evaluation}
\label{sec:evaluation}

This section evaluates GMM's performance implications for CPU core MM
and XPU MM. For CPU MM, the experiments focus on both
single-threaded performance and scalability of standard benchmarks.
For XPU MM, two GMM-based lightweight MM systems are evaluated for
Nvidia and Huawei XPU cards, respectively. The XPU MM experiments 
focus on bandwidth of data transfers and XPU page faults. 

\subsection{Evaluation Setup}

For CPU MM, the baseline system is Linux-6.6. Two systems are 
compared, including GMM and GMM-XPUVisible. They differ by whether 
the \texttt{mmap()} system calls are modified to use 
\texttt{MAP\_XPU\_VISIBLE} which maintains the logical page table for 
both the CPU and XPUs.  
All CPU workloads use \texttt{jemalloc()} v5.3.0. 
CPU MM systems are evaluated on a dual-socket AMD EPYC 7742 server 
(128 cores) with 512 GB 3200 MHz DDR4. 

For XPU MM, each vendor provides a basic memory management system. 
Both of the basic MM systems do not support unified virtual address  
space.
Nvidia provides two extra GPU memory management systems supporting 
unified virtual address space like all GMM-based XPU drivers, 
including the Nvidia UVM driver and the Nvidia HMM driver.  
GMM-based XPU MM drivers and Nvidia UVM driver additionally support 
XPU memory oversubscription.

XPU MM systems are evaluated on two machines: a Nvidia A100 PCIe 3.0 
system hosted by a dual socket Intel Xeon Gold 6240 with 512 GB 3200 
MHz DDR4, and a Huawei 910B PCIe 4.0 system hosted by a quad-socket 
Kunpeng 920 5250 with 1.5 TB 3200 MHz DDR4.

\subsection{XPU MM Performance }

\begin{figure*}[t!]
    \centering
    \includegraphics[width=0.95\textwidth]{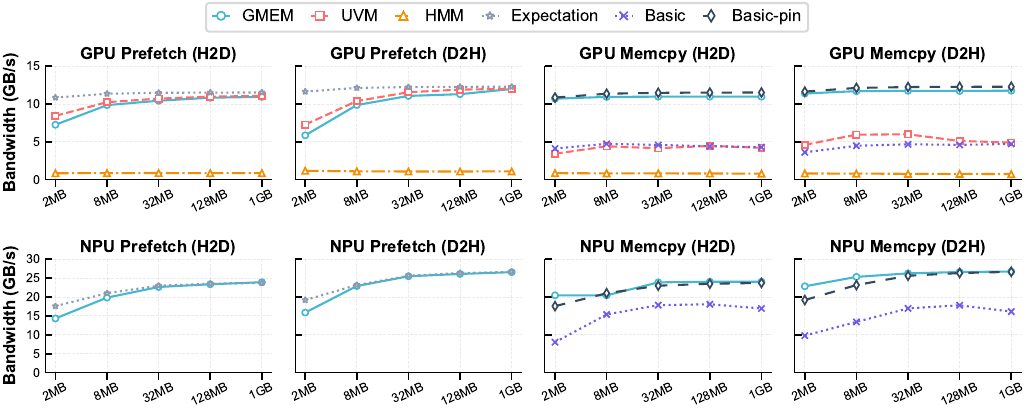}
    \caption{Data Transfer Bandwidth Comparisons on A100-PCIe3 and 910B-PCIe4}
    \label{fig:xpu-bw}
\end{figure*}

\begin{table}[t]
\begin{tabular}{lccc}
\toprule
            & NV-basic & NV-UVM & NV-GMM \\
\midrule
1st Iter    & 0.905s   & 3.387s & 3.319s \\
Avg of 1K Iters & 0.519s   & 0.522s & 0.520s \\
\bottomrule
\end{tabular}
\caption{Training latency of llama2-70B on single A100-PCIe3.0}
\label{tab:gpu-training}
\end{table}

\begin{figure}[t]
    \centering
    \includegraphics[width=0.45\textwidth]{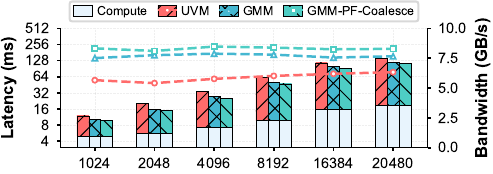}
    \caption{GPU fault-driven migration bandwidth with {M,N,K} float16
    matrix multiplications. M=N=10,240 and the x-axis denotes the K value.}
    \label{fig:gpu-pf}
\end{figure}

GMM significantly improves bandwidth of data prefetching, buffer 
copying and XPU page faults, while matching vendor behavior for real 
applications. 

\paragraph{Real-world workloads.}

Two comprehensive training workloads were evaluated: 

\begin{itemize}
    \item Training LLaMA-2 70B with PyTorch Transformers on A100 
    shows that UVM 
    and GMM incur a higher first-iteration cost due to GPU page 
    faults, but achieve identical long-run training throughput (see 
    Table \ref{tab:gpu-training}.
    \item Training LLaMA-7B (10-layer) with MindSpore on 16 NPUs 
    shows GMM achieves within 0.5\% of the vendor NPU MM driver, 
    indicating no penalty for sitting in the steady state. 
\end{itemize}

\paragraph{CPU - XPU Interaction Bandwidth.}

Figure \ref{fig:xpu-bw} compares bidirectional bandwidth of 
unified-pointer prefetching and pointer-to-pointer memcpy across 
various chunk sizes on both GPUs and NPUs. 

\textbf{GPU results (A100).} 
\textit{Prefetch: }
UVM achieves higher bandwidth than GMM because the GMM-based GPU 
driver reuses the UVM's higher-level page-table operations. 
The bandwidth gap is diminishing as the chunk size grows due to 
GMM's internal workqueue parallelism. HMM constantly obtains around 
1GB/s bandwidth because it migrates pages at 4KB granularity.
\textit{Memcpy: } GMM consistently outperforms UVM, HMM and the 
vendor basic driver approaching the theoretical bandwidth achieved 
by pinned memory, as no coordination penalty between two MM systems 
exist in GMM system. HMM consistently achieves around 1GB/s 
bandwidth due to 4KB page granularity.

\textbf{NPU results (910B).} 
\textit{Prefetch: }
GMM achieves similarly data prefetching throughput as obtained by 
pinned memory, due to the extra cost of control-plane and page table 
manipulations. The bandwidth overlaps as chunk size increases, where 
GMM's internal workqueue hides the extra cost.
\textit{Memcpy: }
GMM is closely tracking basic-pin but without requiring pinned CPU 
memory. The basic driver suffers from coordination penalty as well 
as the GPU basic and UVM drivers.

\paragraph{Fault-Driven GPU Data Migration.}

Figure \ref{fig:gpu-pf} evaluates page-fault bandwidth using FP16 
matrix multiplication. The benchmark evaluates a FP16 matrix 
multiplication GPU kernel {C=A*B}, with fixed N and M values of 
10,240, varying the k from 1,024 to 20,480. 
The A and B matrices are prepared on CPU, and C is zero-filled on 
GPU,
so the end-to-end time includes the computation and implicit
data transfers of A and B from the host to the GPU.
In summary (1) GMM obtains 30.8\% higher migration bandwidth than 
UVM; (2) GMM-PF-Coalesce further improves this to 41.2\%; 
(3) Instrumentation shows 40\% fewer GPU faults due to GMM's fixed 
2MB translation size, while UVM considers 4KB, 64KB and 2MB page 
sizes.

\paragraph{Memory Oversubscription.} 

\begin{figure}[t]
    \centering
    \includegraphics[width=0.45\textwidth]{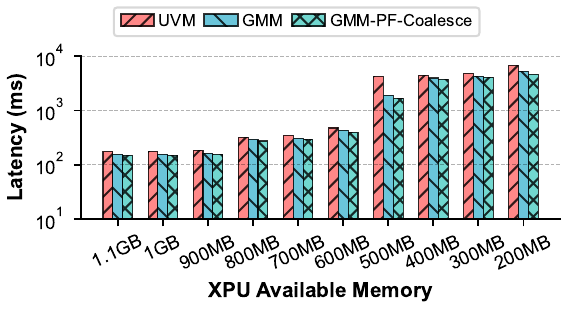}
    \caption{Latency of A100 FP16 matrix multiplication of [10240, 20480, 10240] (costing 1GB memory) with 1GB free XPU memory.}
    \label{fig:pf-latency-bw}
\end{figure}

\begin{figure}[t]
    \centering
    \includegraphics[width=0.45\textwidth]{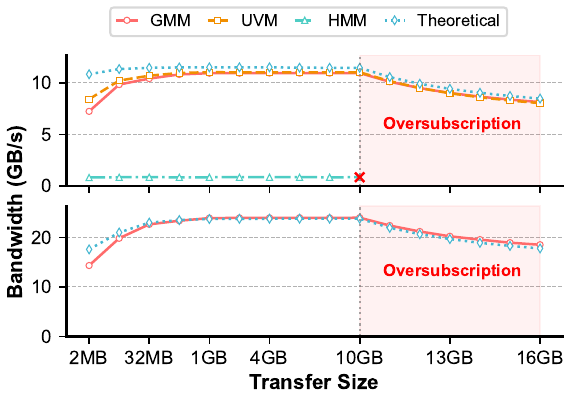}
    \caption{Data prefetching bandwidth with 10GB free XPU memory. Up: A100-PCIe3.0, down: 910B-PCIe4.0}
    \label{fig:prefetch-bandwidth}
\end{figure}

Figure~\ref{fig:pf-latency-bw} and Figure~\ref{fig:prefetch-bandwidth}
restrict XPU memory to investigate memory 
oversubscription performance.

\textbf{Page fault throughput.} GMM and GMM-PF-Coalesce constantly 
outperform UVM by over 13\% for a fixed size of FP16 matrix 
multiplication which costs 1GB memory. GMM-PF-Coalesce brings an 
extra 5\% to 29\% throughput improvement over GMM.

\textbf{Data Prefetching.} On both cards, GMM overlaps with UVM when 
memory is oversubscribed as well as the theoretical throughput 
approximated by basic-pin throughput values. HMM fails to oversubscribe XPU memory.

\subsection{CPU MM Performance}

\begin{figure}[t]
    \centering
    \includegraphics[width=0.45\textwidth]{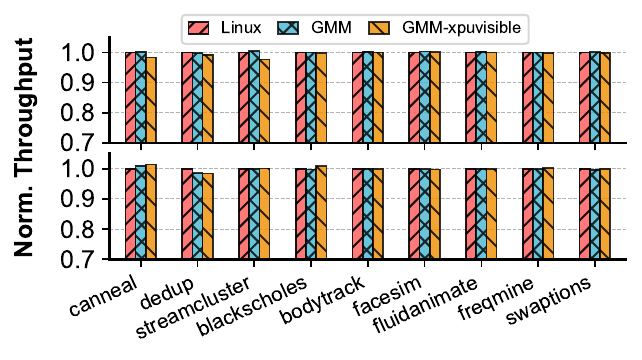}
    \caption{PARSEC performance. Up: single thread, down: eight thread.}
    \label{fig:parsec}
\end{figure}

\begin{figure}[t]
    \centering
    \includegraphics[width=0.45\textwidth]{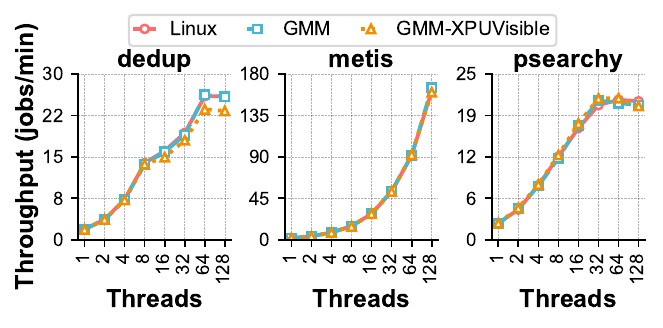}
    \caption{Scalability of memory intensive benchmarks}
    \label{fig:scalable}
\end{figure}

The core MM performance is evaluated on PARSEC 3.0 benchmark suite~\cite{parsec3},
metis and psearchy from MOS benchmark suite~\cite{mosbench}. 
The PARSEC-dedup benchmark, Metis and psearchy particularly stress 
the core MM system.

Figure~\ref{fig:parsec} demonstrates that GMM's throughput differs 
from Linux within 0.5\% in most workloads. 
GMM-XPUVisble introduces a 1.6\% and 2.3\% performance degradation
for single-thread canneal and streamcluster, and 1.6\%
performance degradation for 8-thread dedup, due to maintaining 
logical page tables.

Figure \ref{fig:scalable} demonstrates that both GMM and GMM-XPUVisible mirror a similar scalability pattern with Linux.

\section{Industrial Case Studies}
\label{subsec:industry}

\begin{figure}[t]
    \centering
    \includegraphics[width=0.45\textwidth]{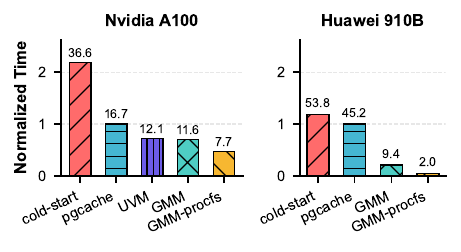}
    \caption{LLM serving switch latency of two models that cannot fit in memory.}
    \label{fig:dual-inference}
\end{figure}

\begin{figure}[t]
    \centering
    \includegraphics[width=0.45\textwidth]{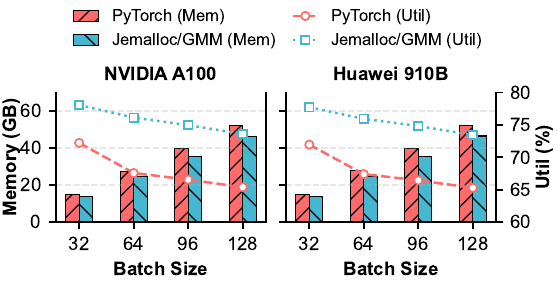}
    \caption{Memory Allocation on OPT-1.3B Training.}
    \label{fig:malloc-GPU-util}
\end{figure}

\begin{figure}[t]
    \centering
    \includegraphics[width=0.45\textwidth]{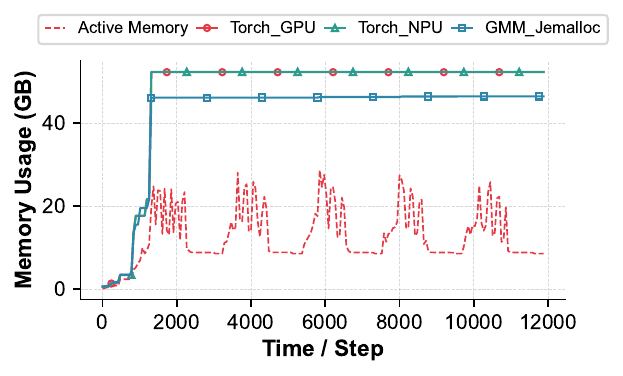}
    \caption{Memory Trace of 1.3B for both GPU and NPU}
    \label{fig:malloc-trace}
\end{figure}

This section performs case studies stemmed from industrial XPU 
deployment experience, including multi-instance LLM serving switch 
and a universal XPU \texttt{malloc()} for low fragmentation. 

\subsection{Multi-instance Serving Switch}

As LLMs are popular, data centers deploy inference service at scale 
and constantly recycle XPUs to serve different inference models
on demand, as the patterns of online requests vary. 
This requires fast serving switch time among multiple inference 
instances.
A naive method is to kill the old inference instance and starts the 
new one to be served. 
However, XPU memory oversubscription allows multiple serving 
processes to be cached, enabling fast serving switch time.

Figure~\ref{fig:dual-inference} compares the serving switch time. 
The A100 card serves two 32B BF16 LLM instances, which add 
up to a total of 130GB memory that exceeds its 80GB memory capacity. 
The 910B card is used to serve two 32B Q8 
LLM instances, which add up to over 65GB memory that exceeds the 
60GB available memory of the 910B card. 
The vanilla method calculates the total time of killing the current 
serving model and starting the new one. 
The cold-start columns include the time to read LLM parameters from 
disk file, while the pgcache columns assume the parameters have been 
cached in the OS page cache. 

\textbf{Serving switch speed.} Normalized to the ideal vanilla case, 
both UVM and GMM achieve 38\% and 44\% faster serving switch speed 
on the GPU case. GMM-procfs is further improved to 2.17x speed up.
For NPU case, the LLM weights are converted to a vendor-specific 
format. 
Therefore, GMM brings a 4.8x speed up by skipping the format 
conversion process, with further improved to a 22.6x speedup 
achieved by GMM-procfs.

\subsection{A Universal \texttt{Jemalloc()} for XPUs}

There are two challenges of dynamic memory allocation for XPUs. 
First, the memory allocation should try to overlap with the XPU 
computation. Second, the fragmentation ratio should be lower since 
XPUs have limited memory capacity. For example,
training features including zero-offload, activation 
recomputation and LoRA can introduce frequent memory allocation and 
deallocation which can cause significant memory fragmentation if a 
memory allocator is not implemented. Deep learning frameworks like 
PyTorch implements a caching allocator which reserves the XPU memory 
via the basic XPU MM API to make sub-allocations. These 
sub-allocations are effectively reinventing a XPU \texttt{malloc()} 
library for fast memory allocation and deallocation.

\textbf{Memory utilization.} Figure~\ref{fig:malloc-GPU-util} 
compares the XPU memory consumption 
using GMM-based \texttt{jemalloc()} library to perform a four-card 
LLM training of OPT-1.3B using DeepSpeed~\cite{guo2024gmlake}. 
GMM-based \texttt{jemalloc()} constantly saves 10\% XPU memory 
compared with PyTorch GPU caching allocator and NPU caching 
allocator, which exhibit identical results. 

\textbf{A100 performance.} The PyTorch native memory allocator that 
directly calls Nvidia basic MM driver API suffers a 87\% training 
throughput decrease compared with PyTorch CachingAllocator. 
Replacing the native memory allocator with GMM-based 
\texttt{jemalloc()} brings around 1\% performance overhead across 
all batch sizes, as some CPU tensors are not extensively replaced by 
GMM's unified pointers.

\textbf{Evidence of reinvention.} Figure~\ref{fig:malloc-trace} further compares the memory 
traces of batch size 128 between PyTorch-GPU, PyTorch-NPU and 
GMM-based jemalloc. The active and reserved memory curves of both 
XPU vendors overlap, providing the evidence of reinvention.
\section{Discussions}

GMM's effectiveness of delegating XPU memory management has been 
empirically studied on two representative XPU vendors. It is worth 
integrating other XPU vendors, e.g. AMD and Intel for preparation of future reviews by the Linux community. We notice a few aspects to be discussed through our experience.

\paragraph{Insights for XPU hardware design}

Although different XPU manufactures adopt different customized 
circuits for domain-specific acceleration, the MMU designs can be 
sharable. It is worth advocating a standardized XPU page table 
format as CPU ISAs, which can be shared by different XPU vendors.

\paragraph{Cache-coherent links}

Existing cache-coherent interconnects are promoted by certain XPU 
manufactures. It is unlikely that these interconnects will become a 
similarly standard as PCIe which is uniformly accepted by CPUs. 
However, GMM can benefit from cache-coherent interconnects. The 
control-plane can be accelerated without sending DMA messages, and 
more advanced memory features like remote mapping can be 
investigated for workloads feature sparse memory access patterns.

\paragraph{Multi-vendor XPU Pooling}

The biggest potential of GMM architecture is to support a unified 
memory resource pooling of different XPU vendors, since the core MM 
takes charge of all GMM-based XPU drivers. Currently, most XPU 
applications are programmed in a distributed manner, where each CPU 
process is bound to a single XPU. While GMM can be easily extended 
to support multiple XPUs (including different generations or 
different vendors), these application scenarios deserve future 
investigation.

\section{Related Work}
\label{sec:related_work}

XPUs employ domain-specific micro-architectural instructions to achieve constant performance optimizations, complementing the general purpose CPUs~\cite{norrie2021design, jouppi2017datacenter, nickolls2010gpu, monmasson2007fpga, farooq2012fpga, kuon2008fpga, dally2020domain, kaplan2024intel, rico2024amd, pomianowski2025amd, smith2025amd, liao2021ascend}.
This raises demand for unified programming~\cite{xsched-ref34, xsched-ref47, xsched-ref62}. Furthermore, there exists demand for unified virtual address space to alleviate programmers' burden, resulting in software solutions for unified virtual memory~\cite{allen2021depth, shao2022oversubscribing, ganguly2019interplay, choi2022memory, allen2024fine, long2023deep, pratheek2024suv, zhu2022uvm}.
Some hardware manufacturers support cache-coherent interconnects between CPUs and XPUs, allowing their MMUs to work on the same physical address space~\cite{emt-ref71, emt-ref96, arsid2025ultra, wei20239, antepara2025roofline, weisgut2150cxl, foley2017ultra, hao2017supporting}.

Commercial CPUs and XPUs mostly utilize multi-level radix tree page tables~\cite{emt-ref11, emt-ref17, emt-ref19, nvidia-gp100-mmu}, causing their memory systems employ software abstractions to hide hardware differences.
The OS core MM systems have been well studied~\cite{cortenmm-ref9, cortenmm-ref10, cortenmm-ref12, cortenmm-ref13, cortenmm-ref15, cortenmm-ref16, cortenmm-ref31, cortenmm-ref33, cortenmm-ref40, cortenmm-ref43, cortenmm-ref65, zhang2025cortenmm, chai2025emt
}. They all utilize software abstractions to hide the differences between CPU ISAs. Recent work mostly focuses on advanced memory features, e.g. huge pages~\cite{emt-ref62, emt-ref69, emt-ref81, emt-ref84, emt-ref85, zhu2020comprehensive} and memory tiering~\cite{emt-ref37, emt-ref61, emt-ref71, emt-ref78, emt-ref92, emt-ref103
}. Furthermore, application-level malloc libraries are studied to enhance the programmability and memory utilization~\cite{dice2002mostly, evans2006scalable, hunter2021beyond, hudson2006mcrt, leijen2019mimalloc, yang2023numalloc, tcmalloc_overview}.
However, XPU MMs are bound with XPU manufacturers, causing a proliferation of reinvented memory systems. Recent work mostly focuses on fixing the XPU MMs at the application level~\cite{
wang2018superneurons, shriram2019dynamic, peng2020capuchin, guo2024gmlake, devito_cuda_caching_2022}. As more XPU manufacturers are involved, these solutions will demand reinvention on other XPU MMs as well.

\section{Conclusion}
\label{sec:conclusion}


GMM redesigns the core MM by adding an explicit logical mapping 
abstraction that unifies XPUs in the OS core memory management path. 
Our work shows the problem of independent XPU memory management 
which is currently leading to fragmented memory-related software. 
Practices of implementing lightweight GMM-based XPU drivers 
validate 
GMM's design principle which allows ease of XPU MM delegation.
GMM eliminates the coordination penalty between the core MM and 
independent proprietary XPU MM drivers and achieves competitive data 
transfer bandwidth on both representative vendors. 
Additionally, this work exercises two case studies stemmed from real 
industrial XPU demands, demonstrating its effectiveness of 
leveraging advanced memory features.
As a straightforward design, GMM allows the system community to 
simultaneously investigate memory mechanisms and policies on a 
significantly broader combinations of heterogeneous machines.
\section{Acknowledgments}
\label{section:ack}

The original idea of this work was a proof-of-concept demonstration based on FreeBSD and open-sourced MMU drivers including Intel's integrated GPU and IOMMU~\cite{zhu2023gmem}. This work extended the original idea and exercised it on Linux with commercial AI chips for real-world workloads, evaluating the effectiveness of operating system memory management subsystems on AI accelerators. We thank the openEuler developers Tianyuan Bi, Jun Chen,
Yuhang Chen, Liming Huang, Chao Liu, Wupeng Ma, Cunshu Ni,
Tianchen Xiong, Qing Xiu, Huijie Xu, Junming Xu, and
Yanchao Yang for their help with the implementation.


\clearpage
\bibliographystyle{plain}
\bibliography{refs}

\end{document}